\documentstyle[12pt,psfig]{article}

\newcommand{\ra}{\rightarrow}

\newcommand{\be}{\begin{equation}}
\newcommand{\ee}{\end{equation}}

\newcommand{\bea}{\begin{eqnarray}}
\newcommand{\eea}{\end{eqnarray}}
\newcommand{\bef}{\begin{figure}}
\newcommand{\eef}{\end{figure}}

\begin{document}
\vskip 0.2in
\begin{center}

{\large{\bf Charmonium at LHC: Production, Propagation and Decay}}
\end{center}
\vskip 0.1in
\begin{center}
{\normalsize 
Pradip Roy$^a$\footnote{Part of the work has been done
at CERN Theory Division, CERN, Geneva}, Abhee K. Dutt-Mazumder$^a$, 
Jane Alam$^b$, and Bikash Sinha$^{a,b}$
}
\vskip .1in
a) Saha Institute of Nuclear Physics, 1/AF Bidhan Nagar, Kolkata, India
\\
b) Variable Energy Cyclotron Centre, 1/AF Bidhan Nagar, Kolkata, India
\\
\vskip .1in
\end{center}

\addtolength{\baselineskip}{0.4\baselineskip} 

\date{today}
\parindent=20pt
\vskip 0.1in
\begin{abstract}
The productions of muon pairs from the decay of heavy quarkonia have been
evaluated for different centrality of the nuclear collisions at LHC energies.
The effects of the various comover scenarios on 
the survival probability of the heavy quarkonia have
been considered.  The effects of shadowing and comover  
suppressions on the dilepton spectra originating from the decays
of $J/\psi$ is found to be substantial.
The dilepton yield from the  thermal $J/\psi$ has also been
estimated and found to be small.

\end{abstract}
\section*{I. Introduction}
Ever since the possibility of creating quark gluon plasma (QGP)
in relativistic heavy ion collision was envisaged, numerous 
signals were proposed to probe the properties of such an exotic
state of matter.  In this context 
Satz and Matsui~\cite{satz} had suggested that
the production of heavy quark resonances ($J/\psi$) will be suppressed 
as a result of colour Debye screening in a hot
and dense system of quarks, anti-quarks and gluons. 
This suppression could
be detected experimentally through the dileptonic decay mode of these
resonances. ALICE dimuon spectrometer~\cite{muonlab} is dedicated to look 
for this
type of signal. However, it is a daunting task to 
disentangle  the contributions of the heavy quarkonium states
to muon spectrum due to the 
background from several other sources, {\it e.g.}
Drell-Yan, semileptonic
decay of open heavy flavoured mesons ($D{\bar D}, B{\bar B}$) etc. 
Low energy muons
from kaons and pions also constitute a large background. 


In this work we shall estimate the dimuon  production from the decay
of both hard  ({\it i.e.} $J/\psi$ 
produced from initial hard process, will
be called hard $J/\psi$ hereafter) and thermal $J/\psi$'s.
In heavy ion collisions the production and  decay of hard $J/\psi$ or
$\Upsilon$ proceeds 
through the following three steps:
(i) the production of pair of heavy quarks (perturbative), 
(ii) their resonance interactions to form the bound state (non-perturbative)
and (iii) the propagation of 
the quarkonia through the  medium and their subsequent 
decay to dileptonic modes with certain branching ratios.

The initial state in relativistic heavy ion collisions 
consists of either hadronic matter  
or QGP depending on the incident energies
of the colliding nuclei. At LHC energies
the formation of QGP is almost unavoidable.
Even if the system is formed in QGP phase it
will revert back to hadronic phase due to the cooling
of the expanding QGP system
and hence the interaction of the $J/\psi$ formed
in initial hard collision with the hadronic
matter is inevitable. Therefore,, we need
to consider the survival probability of those
$J/\psi$ due to its interactions with the hadronic
medium.

At high temperatures a non-negligible 
number of $J/\psi$ mesons is expected in the thermal medium, 
which decay to lepton pair. The dilepton spectra 
originating from these thermal $J/\psi$ is 
also estimated here.

The paper is organized as follows.
In section II we shall describe the formalism for the production, propagation
and decay of hard and thermal $J/\psi$'s.
In section III results of our calculations will
be presented followed by summary and discussion in section IV.

\section*{II. Production, Propagation and Decay of Hard and Thermal $J/\psi$}

\subsection*{a. Production}

In this section we shall consider the hard $J/\psi$ production in the colour
evaporation model (CEM)~\cite{cem} and their decays to lepton pairs.
As mentioned before this consists of two stages:
(i) production of a $c{\bar c}$ pair (perturbative process) and (ii) 
subsequent non-perturbative evolution into asymptotic states.  
We have considered those
hard processes which can contribute to $c\bar{c}$ productions irrespective
of their colour and spin-parity. The colour 
neutralization occurs by the interactions (one or more soft gluon emission)
with the surrounding colour fields and this step is considered to be 
non-perturbative. 
In CEM quarkonium production is treated identically to open heavy flavour
production with the exception that in the case of quarkonium, the invariant
mass of the heavy quark pair is restricted below the open meson threshold,
which is twice the mass of the lowest meson mass that can be formed with 
the heavy quark. Depending on the quantum numbers of the initial $Q{\bar Q}$
pair and the final state quarkonium, a different matrix element is needed
for the resonance production. The effects of these non-perturbative matrix
elements are combined into the universal factor $F[nJ^{PC}]$ which is a process
and kinematics independent quantity~\cite{FnJ}. 
It describes the probability that the
$Q{\bar Q}$ pair forms a quarkonium of given spin ($J$), parity ($P$)
and charge conjugation ($C$). The production cross section for a $J/\psi$ or
$\Upsilon$ is therefore given by~\cite{FnJ}
\be
\sigma[R(nJ^{PC})] = F[nJ^{PC}]\,{\tilde {\sigma}}[Q{\bar Q}]
,
\label{eq9}
\ee  
where the non-perturbative (long distance) factor can be written 
in terms of the probability
to have colour singlet state (1/9) and the fraction $\rho_R$ of each specific
charmonium state. The perturbative contribution (short distance) 
is given by
\be
{\tilde{\sigma}}[Q{\bar Q}] = \int_{2m_Q}^{2m_{D/B}}\,dM_{Q{\bar Q}}^2
\,\frac{d\sigma[Q{\bar Q}]}{dM_{Q{\bar Q}}^2}.
\label{eq10}
\ee

The contributions to heavy quark production in leading order come from
$q\,{\bar q}\,\ra\,Q\,{\bar Q}$ and $g\,g\,\ra\,Q\,{\bar Q}$. The differential
cross-section for heavy flavour production in hadron-hadron 
collision is given by~\cite{vogt}
\be
\frac{d\sigma}
{dM_{Q{\bar Q}}^2 dy}
[h_A h_B\ra Q {\bar Q}X] = 
\frac{H(x_a,x_b,Q^2)}{s},
\label{eq11}
\ee
where 
\bea
H(x_a,x_b,Q^2)& =& \sum_f\,\left[\frac{}{}q_f^{h_A}(x_a,Q^2)
{\bar q}_f^{h_B}(x_b.Q^2)+\frac{}{}{\bar q}_f^{h_A}(x_a,Q^2)
q_f^{h_B}(x_b,Q^2){\hat {\sigma}}_{q{\bar q}\ra Q{\bar Q}}\right]\nonumber\\
&&+g^{h_A}(x_a,Q^2)g^{h_B}(x_b,Q^2){\hat {\sigma}}_{gg\ra Q{\bar Q}}.
\label{eq12}
\eea
$x_{a,b} = M\,e^{\pm y}/\sqrt{s}$, $\sqrt{s}$ being the centre of mass
energy of the hadronic system. $q_f$'s and $g$'s are the parton distribution
functions (PDF) to be taken from CTEQ, MRST or GRV~\cite{pdf}. 
Combining eqs.(\ref{eq9}),(\ref{eq10}) and (\ref{eq11}) we obtain the 
cross-section for resonance production  per unit rapidity as,
\be
\frac{d\sigma}
{dy}
[h_A h_B\ra R X] = 
\frac{\rho_R}{9}\,\int_{2m_Q}^{2m_{D/B}}\,dM_{Q{\bar Q}}^2
\frac{H(x_a,x_b,M_{Q{\bar Q}}^2)}{s}.
\label{eq13}
\ee
We take $\rho_R$ = 0.5 (0.207) for $J/\psi$ ($\Upsilon$).

Next we consider $J/\psi (\Upsilon)$ production in $p-A$ and $A-B$ collisions.
To this end we first briefly mention the necessary formulae in Glauber 
model~\cite{glau,wong}
The total inelastic cross-section
in $A-B$ collisions at an impact parameter $b$ is given by
\be
\frac{d\sigma_{\mathrm {in}}^{AB}}{d\vec{b}} = 1-\left[\frac{}{}
1-T_{AB}(\vec{b})\,\sigma_{\mathrm{in}}^{NN}\right]^{AB}
\equiv\,1 - P_0(b),
\label{eq18}
\ee
where $T_{AB}(b)$ is the nuclear overlap function given by
\be
T_{AB}(b) = \int\,d^2s\,T_A(b)\,T_B(|\vec{b}-\vec{s}|)
\label{eq19}
\ee
The nuclear thickness functions are normalized to unity, {\it{i. e.}}
$\int\,d^2b\,T_A(b) = \int\,d^2b\,dz\,\rho_A(b,z) = 1$

Generally we are interested in the cross-sections for a set of
events in a given centrality range defined by the trigger settings.
Centrality selection corresponds to a cut on the impact parameter
$b$ of the collisions. The sample of events in a given centrality
range $0 \leq b \leq b_m$, contains a fraction of the 
total inelastic cross-section.
This fraction is defined by
\bea
f(b_m)&=&\frac{
\int_0^{b_m}d{\vec b}\,
\frac{d\sigma_{\mathrm {in}}^{AB}}{d\vec b}}
{\int_0^\infty\,d\vec{b}\,
\frac{d\sigma_{\mathrm{in}}^{AB}}{d\vec{b}}}
\label{eq20}
\eea

Now we discuss the $J/\psi$ survival probability when it propagates 
through the hot hadronic medium 
produced in nucleus-nucleus collisions.
After creation the $J/\psi$ meson can interact with other nucleons in the
target and the projectile and may get destroyed mainly due to $J/\psi - N$
interactions. The cross-section for $J/\psi (\Upsilon)$ production in
$p-A$ collisions can be written as
\be
\sigma_{J/\psi}^{pA}(b_m) = A\int_0^{b_m}\,d{\vec b}\,dz\,\rho({\vec b},z)\,
\exp{\left[-(A-1)\int_z^\infty\,\sigma_{\mathrm{abs}}\,\rho({\vec b},{z^{\prime}})\,
d{z^{\prime}}\right]}\,{{\sigma}}_{J/\psi}^{NN}, 
\label{eq20a}
\ee
where ${\bar{\sigma}}_{J/\psi}^{NN}$ is obtained from 
eq.(\ref{eq13}). The interpretation of 
the above equation is as follows.
The resonance is formed at ${\vec r} = ({\vec b},z)$ where the density 
of the target nucleus is $\rho({\vec r})$. It can travel in forward
direction ($z$) at constant impact parameter and its intensity is
attenuated due to $J/\psi-N$ inelastic collisions. The exponential
factor accounts for this attenuation loss. 

The generalization of eq.(\ref{eq20a}) in nucleus-nucleus collisions is 
straightforward.
The $J/\psi (\Upsilon)$ production cross-section in $A-B$ collisions in
the impact parameter range $0 \leq b \leq b_m$ can be written as 
\bea
\frac{d\sigma_{J/\psi(\Upsilon)}^{AB}}{d^2bdy}(b)& = &
\frac{d{{\sigma}}_{J/\psi(\Upsilon)}^{NN}}{dy}\,
AB\,\int\,d{\vec s}\,dz_1\,dz_2\,\rho_A(s,z_1)\,
\rho_B(|{\vec b}-{\vec s}|,z_2)\nonumber\\
&&\times\,\exp{\left(-(A-1)\int_{z_1}^{\infty}\,\sigma_{\mathrm{abs}}\,
\rho_A({\vec s},{z^{\prime}})\,
d{z^{\prime}}\right)}\,\nonumber\\
&&\times\,\exp{\left(-(B-1)\int_{z_2}^{\infty}\,\sigma_{\mathrm{abs}}\,
\rho_B(|{\vec b}-{\vec s}|,{z^{\prime}})\,
d{z^{\prime}}\right)}
\label{eq21}
\eea
where ${{\sigma}}_{J/\psi(\Upsilon)}^{NN}$ is obtained using the 
nuclear parton distribution functions (PDF) containing shadowing effects. 

\subsection*{b. Propagation}

The experimental data suggest that besides the absorption due 
to $J/\psi$-nucleon interactions, there are additional sources
of $J/\psi$ absorption in nuclear collisions. The produced 
$J/\psi$ particles can interact with the relatively heavier 
mesons (such as $\rho$, $\eta$ etc.) expected to be produced 
at a proper time $\tau_0$. Such interactions can lead to the
disappearance of $J/\psi$. Not all the hadrons produced in nucleus-nucleus
collisions (co)move with the $J/\psi$. So the number density of 
comoving hadrons is given by $n_{\mathrm co} = f_c\,n_h$,
where $n_h$ is density of produced hadrons and $f_c$ is the fraction
that (co)moves with the $J/\psi$ . In the following sections
we shall discuss three different `comoving' scenarios. 

\subsubsection *{Scenario I } 
In this section we outline the comoving scenario of Ref.~\cite{wong}.
We consider that a hadronic matter
is formed at a proper time $\tau_0$ with density $n_h (\tau_0)$ and
evolves hydrodynamically. If we assume Bjorken's scaling~\cite{bj} solution,
then at a later time ($\tau$) the density of hadron ($n_h(\tau)$) is given by
\be
n_h(\tau) = \frac{n_h(\tau_0)\,\tau_0}{\tau}
\label{eq21a}
\ee
During the time span from $\tau_0$ to $\tau_F$ (the freeze-out time) the
$J/\psi$ particles interact with the comoving fluid. The survival probability
$S_{\mathrm co}$ is given by~\cite{npbv}
\be
S_{\mathrm co} = \exp[{-\int_{\tau_0}^{\tau_F}\,d\tau \langle\,
\sigma_{\mathrm co}\,v_{\mathrm rel}\,f_c\,n_h(\tau)\rangle}]
\label{eq21b}
\ee
Using eqs.(\ref{eq21a}) and (\ref{eq21b}) we get
\be
S_{\mathrm co} = \exp[-\langle\,
\sigma_{\mathrm co}\,v_{\mathrm rel}\,\rangle\,\tau_0\,f_c\,n_h(\tau_0)\,
\ln\frac{\tau_F}{\tau_0}],
\label{eq21c}
\ee
where $\sigma_{\mathrm co}$ is the $J/\psi$-comover cross-section 
leading to the breakup of $J/\psi$ particle, $v_{\mathrm rel}$ is the
relative velocity between the $J/\psi$ and the comover. 
We have used a constant value
for $\sigma_{\mathrm co}$ in our calculation. 


The hadron density  $n_h$ can be related to the multiplicity of produced
hadrons at $\tau_0$ from the row-on-row collisions of the projectile and
the target nucleons. The volume of the row is 
$\sigma_{\mathrm in}^{\mathrm NN}\,\tau_0\,dy$ at proper time $\tau_0$.
The hadron density of this row is therefore:
\be
n_h(\tau_0,{\vec b},{\vec s}) = \frac{1}{\sigma_{\mathrm in}^{\mathrm NN}\,
\tau_0}\,\frac{dN_{AB}}{dy}({\vec b},{\vec s})
,
\label{eq21d}
\ee
where $dN_{AB}/dy({\vec b},{\vec s})$ is the hadron rapidity
density which can be calculated from the participant density,
$n_w^\prime({\vec b},{\vec s})$. For a row-on-row collision,
it is given by
\be
n_w^\prime({\vec b},{\vec s}) = \sigma_{\mathrm in}^{\mathrm NN}\,\left[
A\,T_A({\vec b}) + B\,T_B({\vec b}-{\vec s})\right]
,
\label{eq21e}
\ee
and the multiplicity is obtained as
\be
\frac{dN_{AB}}{dy}({\vec b},{\vec s}) \sim \frac{1}{2}\,\frac{dN_{NN}}{dy}\,
 \sigma_{\mathrm in}^{\mathrm NN}\,\left[
A\,T_A({\vec b}) + B\,T_B({\vec b}-{\vec s})\right]
,
\label{eq21f}
\ee
where $\frac{dN_{NN}}{dy}$ is the multiplicity in nucleon-nucleon collisions.
Combining eqs.(\ref{eq21b})-(\ref{eq21f}) we obtain the
$J/\psi$ survival probability, $S_{\mathrm co}$ in a comoving scenario
as
\be
S_{\mathrm co}({\vec b},{\vec s}) = \exp{\left(-c_0\,\left[
A\,T_A({\vec b}) + B\,T_B({\vec b}-{\vec s})\right]\right)
},
\label{eq21g}
\ee
where $c_0$ is given by
\be
c_0 = \langle \sigma_{\mathrm co}\,v_{\mathrm rel}\,\rangle\,
\ln\frac{\tau_F}{\tau_0}\,f_c\,\frac{1}{2}\frac{dN_{NN}}{dy}
\label{eq21h}.
\ee
For hadron rapidity density in nucleon-nucleon collision we use
the following form~\cite{nardi}: $dN_{NN}/dy = 
2.5 - 0.25\,\ln(s) +0.023\,\ln^2(s)$.

\subsubsection *{Scenario II}
 
In Ref.~\cite{vogt} it is assumed that all the comoving hadrons
are generated from the participating nucleons, which might be a reasonable
assumption at SPS energies. However, at higher energies it needs
modifications, as discussed below.
One would expect that the total multiplicity
in nucleus-nucleus collisions at RHIC and LHC energies will come from both
soft ($\sim N_{\mathrm part}$) as well as hard ($\sim N_{\mathrm coll}$)
collisions. It has been shown recently in~\cite{nardi} 
that about 10 \% of the total multiplicity at RHIC energies
 comes from hard collisions,
{\i.e.} the total hadron multiplicity in nucleus-nucleus collisions at
a given $b$ can be written as
\be
\frac{dN_{AB}}{dy}(b) = \left[(1-x)\,\frac{N_{\mathrm part}
(b)}{2}
+ x\,N_{\mathrm coll}(b)\right]\,\frac{dN_{NN}}{dy},
\label{eq21i}
\ee
where 
\bea
N_{\mathrm part}(b)&=&
\int\,d{\vec s}\,n_w({\vec b},{\vec s})\nonumber\\
N_{\mathrm coll}(b)&=& AB\,\sigma_{\mathrm in}^{\mathrm {NN}}
\,T_{AB}({\vec b})
\label{eq21j}
\eea
The participant density $n_w({\vec b},{\vec s})$ in this case is given
by 
\bea
n_w({\vec b},{\vec s}) &=& AT_A({\vec s})\left\{1 - \left[1-
\sigma_{\mathrm in}^{\mathrm {NN}}T_B({\vec b}-{\vec s})\right]^A\right\}
\nonumber\\
&+&BT_B({\vec b}-{\vec s})\left\{1 - \left[1-
\sigma_{\mathrm in}^{\mathrm {NN}}T_A({\vec s})\right]^B\right\}
\label{eq21k}
\eea
The initial density of the comoving hadrons in this scenario is:
\be
n_{\mathrm co}(\tau_0) = n_h(\tau_0)\,f_c = f_c\,\frac{dN_{AB}}{dy}\,
\frac{1}{\pi R_A^2\tau_0}. 
\label{eq21l}
\ee
Thus the comoving survival probability is given by
\be
S_{\mathrm co}(b) = \exp[-\langle\,
\sigma_{\mathrm co}\,v_{\mathrm rel}\,\rangle\,
\frac{dN_{AB}(b)}{dy}\,\frac{f_c}{\pi R_A^2}\,
\ln\frac{\tau_F}{\tau_0}],
\label{eq21m}
\ee

\subsubsection *{Scenario III} 

We assume that the comoving density is proportional to the
final $dN_{AB}/dy$ which is a function of centrality. In this
scenario it is also assumed that the final multiplicity of the
produced hadrons depends on both $\langle N_{\mathrm part} \rangle $
and $\langle N_{\mathrm coll} \rangle $ (two component model). 
In this  scenario $S_{\mathrm co}$ can be 
estimated as~\cite{vogt}
\be
S_{\mathrm co}({\vec b},{\vec s}) = \exp[-\langle\,
\sigma_{\mathrm co}\,v_{\mathrm rel}\,\rangle\,
n_w({\vec b},{\vec s})\,
\ln\frac{\tau_F}{\tau_0}],
\label{eq21n}
\ee
where it has been assumed that the number of comoving hadrons
is proportional to the number of participants. At low energies
({\it e.g.} SPS) this might be a valid assumption. However, at higher energies
a certain fraction of comoving hadrons may come from the hard collisions
($\propto N_{\mathrm coll}$). To take into account this fact we modify
the above equation to obtain
\be
S_{\mathrm co}({\vec b},{\vec s}) = \exp\left\{-\langle\,
\sigma_{\mathrm co}\,v_{\mathrm rel}\,\rangle\,
\left[(1-x)\,\frac{n_w({\vec b},{\vec s})}{2}+x\,n_c({\vec b},{\vec s})\right]\,
\ln\frac{\tau_F}{\tau_0}\right\},
\label{eq21o}
\ee
where $n_w({\vec b},{\vec s})$ is given by eq.(\ref{eq21k}) and 
$n_c({\vec b},{\vec s}) = AB\,\sigma_{\mathrm in}^{\mathrm NN}\,T_A({\vec b})\,
T_B({\vec b}-{\vec s})$.

In relativistic heavy ion collisions the yield is the relevant quantity rather
than the cross-section. In order to obtain the differential number 
distribution from the cross-section one has to resort to Glauber model
of nucleus-nucleus scattering~\cite{wong}. 
Incorporating the comoving survival probability in eq.(\ref{eq21})
we obtain the $J/\psi$ production cross-section in $A-B$ collisions
as
\bea
\frac{d\sigma_{J/\psi(\Upsilon)}^{AB}}{dy}(b_m)& = &
\frac{d{{\sigma}}_{J/\psi(\Upsilon)}^{NN}}{dy}\,
AB\,\int\,d{\vec s}\,dz_1\,dz_2\,\rho_A(s,z_1)\,
\rho_B(|{\vec b}-{\vec s}|,z_2)
\,S_{\mathrm co}({\vec b},{\vec s})\nonumber\\
&&\times\,\exp{\left(-(A-1)\int_{z_1}^{\infty}\,\sigma_{\mathrm{abs}}\,
\rho_A({\vec s},{z^{\prime}})\,
d{z^{\prime}}\right)}\,\nonumber\\
&&\times\,\exp{\left(-(B-1)\int_{z_2}^{\infty}\,\sigma_{\mathrm{abs}}\,
\rho_B(|{\vec b}-{\vec s}|,{z^{\prime}})\,
d{z^{\prime}}\right)}\nonumber\\
&\equiv&\,AB{\cal F}(b_m)\,
\frac{d{{\sigma}}_{J/\psi(\Upsilon)}^{NN}}{dy}
\label{eq21p}
\eea

The total number of
$J/\psi (\Upsilon)$ produced in $A-B$ collisions can now be written as
\be
\frac{dN_{J/\psi(\Upsilon)}^{AB}}{dy}(b_m) = \frac{AB{\cal F}(b_m)\,
}{\sigma_{\mathrm{in}}^{AB}(b_m)}
\frac{d{\bar {\sigma}}_{J/\psi(\Upsilon)}^{NN}}{dy}\,
\label{eq22q}
\ee

\subsection*{c. Decay of $J/\psi$ to lepton pairs}
To calculate $h_A\,h_B\,\ra\,R\,X\,\ra\,l^+\,l^-\,X$ from 
hard processes we proceed as follows. 
Symbolically we can write
\be
\frac{d\sigma}
{dy}
[h_A h_B\ra R (l^+ l^-) X] = 
\frac{\Gamma_{R\ra l^+l^-}}{\Gamma_R}\frac{d\sigma}
{dy}
[h_A h_B\ra R X] 
\label{eq14}
\ee

After production the quarkonia  propagate in the medium 
before decaying  to lepton pairs with certain branching ratio. 
The finite width of the quarkonia may be taken into  account 
by folding Eq.~\ref{eq14} with the spectral function ($A_R(M)$) 
of the quarkonia,
\be
\frac{d\sigma}{dy}[h_A h_B\ra R (l^+ l^-) X] = 
\int\,dM^2\,\frac{d\sigma}{dy}[h_A h_B\ra R X]\, 
\frac{\Gamma_{R\ra l^+l^-}}{\Gamma_R}\,A_R(M)
\label{eq16}
\ee
where $A_R(M)$ is given by,
\be
A_R(M) = \frac{1}{\pi}\,\frac{M\Gamma_R}{(M^2-m_R^2)^2+M^2\Gamma_R^2},
\label{eq15}
\ee

Using the above set of equations we obtain the differential cross-section
for the lepton pair production from hard heavy quark resonance decay as
\bea
\frac{d\sigma}{dM^2 dy}[h_A h_B\ra R (l^+ l^-) X] &=& 
\frac{1}{\pi}\,\frac{M\Gamma_{R\ra l^+l^-}}{(M^2-m_R^2)^2+M^2\Gamma_R^2}
\nonumber\\
&&\times\,
\frac{\rho_R}{9}\,\int_{2m_Q}^{2m_{D/B}}\,dM_{Q{\bar Q}}^2
\frac{H(x_a,x_b,M_{Q{\bar Q}}^2)}{s},
\label{eq17}
\eea

Now the dilepton yield from hard $ J/\psi (\Upsilon)$ decay in a
given centrality range $0\,\leq\,b\,\leq\,b_m$ can be written as 
\be
\frac{dN}{dM^2dy}[A B\ra R (l^+ l^-) X] = 
\frac{1}{\pi}\,\frac{M\Gamma_{R\ra l^+l^-}}{(M^2-m_R^2)^2+M^2\Gamma_R^2}\,
\frac{dN_{J/\psi(\Upsilon)}^{AB}}{dy}(b_m)
\label{eq22}
\ee

In a similar way one can also calculate the the total number of 
heavy quark pairs,
in a given centrality range. These are:
\bea
N_{Q{\bar Q}}^{AB}(b_m)& =& {\cal R}\,{\bar {\sigma}}_{J/\psi(\Upsilon)}^{NN}
\label{eq25}
\eea
where ${\cal R}$ is given by
\bea
{\cal R}&=&\frac{\langle N_{\mathrm{binary}} \rangle (b_m)}
{\sigma_{\mathrm{in}}^{NN}}\nonumber\\
&=&
\frac{\int_0^{b_m}\,d{\vec b}\,AB\,T_{AB}(b)}
{\int_0^{b_m}\,d{\vec b}\,[1-\{1-T_{AB}(b)\sigma_{\mathrm{in}}^{NN}\}^{AB}]}
\label{eq26}
\eea

\subsection*{d. Thermal $J/\psi$ Decay}
The muon yields from thermal $J/\psi$ decay during the 
life time of the fire ball is expected to be 
small due two reasons: (i) less abundance of the $J/\psi$ in the
thermal system of temperature $\sim$ of few hundred MeV 
and (ii) the probability of the $J/\psi$ to decay within the lifetime
of the hadronic system is a very small due to tiny width of the $J/\psi$. 
However, for the sake of completeness we give the thermal spectra below: 
\bea
\frac{dN}{dM^2dy}& =& \frac{2J+1}{(2\pi)^2}\,\pi\,R_A^2\,3T_i^6\,\tau_i^2\,
M\,\Gamma_{J/\psi\ra\,l^+l^-}\,
\nonumber\\
&&\times\,
\frac{1}{\pi}\left[\frac{M\Gamma_{\mathrm{tot}}}
{(M^2-m_{J/\psi}^2)^2+(M\Gamma_{\mathrm{tot}})^2}\right]
\,\int_{T_f}^{T_i}\,\frac{dT}{T^5}\nonumber\\
&&\times\,\int_{M/T}^\infty\,z\,dz\,\int_{\eta_1}^{\eta_2}\,
\frac{1}{\exp(z\cosh(y-\eta))-1}
\label{eq8a}
\eea
where $z=M_T/T$. 

As mentioned above due to the small width of the $J/\psi$ most
of them will decay after the fireball freeze-out. The yield
from these $J/\psi$ have also been estimated using Cooper-Frye
formula~\cite{cooper} and added to the thermal contributions.
The expression for this contribution is given below:
\bea
\frac{dN}{dM^2dy}& =&\frac{g\,R_A^2\,\tau_F}{4\pi^2}\, 
\frac{1}{\pi}\left[\frac{M\Gamma_{R\ra\,l^+l^-}}
{(M^2-m_{J/\psi}^2)^2+(M\Gamma_{\mathrm{tot}})^2}\right]\nonumber\\
&&\times\,\sum_{n=1}^\infty\,(+)^{n+1}\,\int\,K_1(n\,m_T/T)\,m_T\,d^2p_T
\eea

\section*{III. Results}
The variation of ${\cal F}(b_m)$, which is the ratio of $J/\psi$ production
cross-section in $A-B$ collisions (scaled to $p-p$) to that in 
nucleon-nucleon collision as a function of centrality ($b_m$)
is shown in fig.~(\ref{fig1}). Here we have assumed that there is no
suppression due to comoving hadrons. In presence of comovers ${\cal F}(b_m)$
will decrease further. However, in the evaluation of dilepton yield we
have incorporated the suppression due
to comover in three different scenarios as discussed earlier. 
\bef
\centerline{\psfig{figure=fig1.eps,height=8cm,width=10cm}}
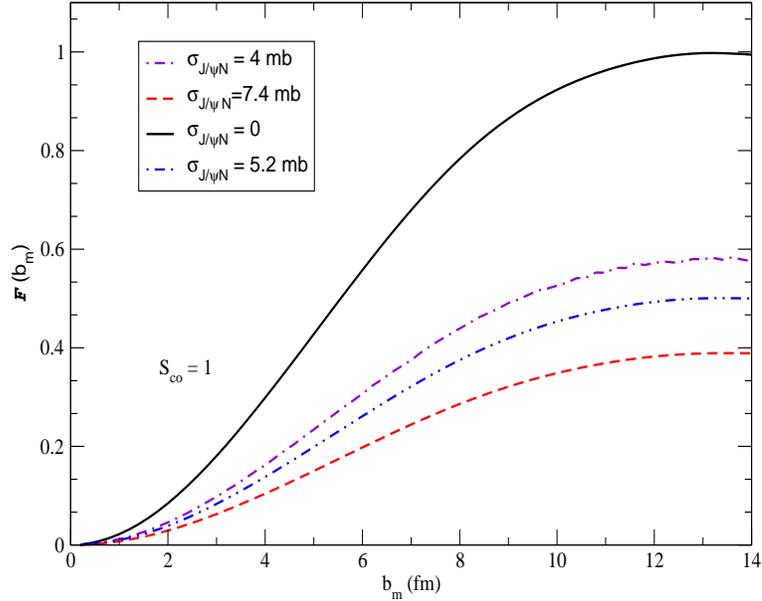
\caption{The centrality dependence of the suppression factor.
}
\label{fig1}
\eef

In fig.~(\ref{fig2}) we show the centrality dependence of the total 
$J/\psi$ production cross-section in nucleus-nucleus collisions. It is
seen that the survival probability is very sensitive to the choice of
$\sigma_{J/\psi N}$ cross-section as well as to the comoving scenario adopted.

\bef
\centerline{\psfig{figure=fig2.eps,height=8cm,width=8cm}}
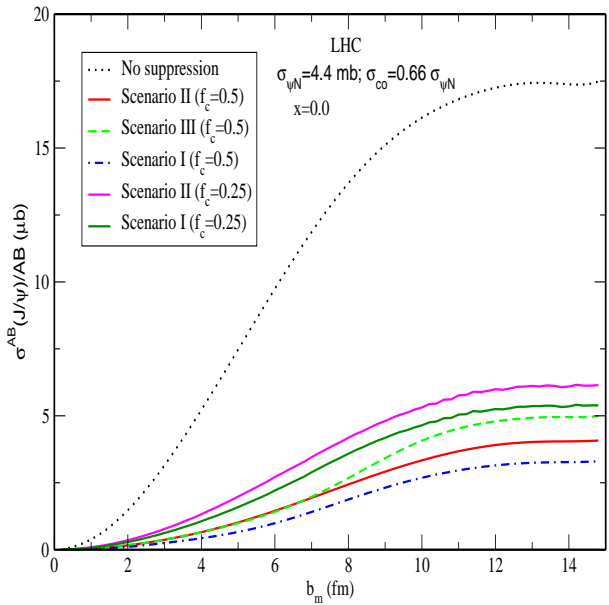
\caption{Total $J/\psi$ cross-section in different comoving scenarios 
(see text) as function of centrality at LHC energies.}
\label{fig2}
\eef

In the second scenario of comoving absorption the survival probability becomes
function of $b$ only. So in fig.~(\ref{fig3}) we plot $S_{\mathrm co}$
as a function of impact parameter. The suppression depends on two
factors: (i) what fraction ($x$) of produced hadrons come from hard processes 
and (ii) what fraction ($f_c$) of these hadrons (co)move with the $J/\psi$. 
To show the sensitivity on these two factors we have chosen various 
combinations of $x$ and $f_c$. It is seen that for a given $f_c$ as
$x$ increases (which is possible as the beam energy increases) the survival
probability goes down.
\bef
\centerline{\psfig{figure=fig3.eps,height=9cm,width=12cm}}
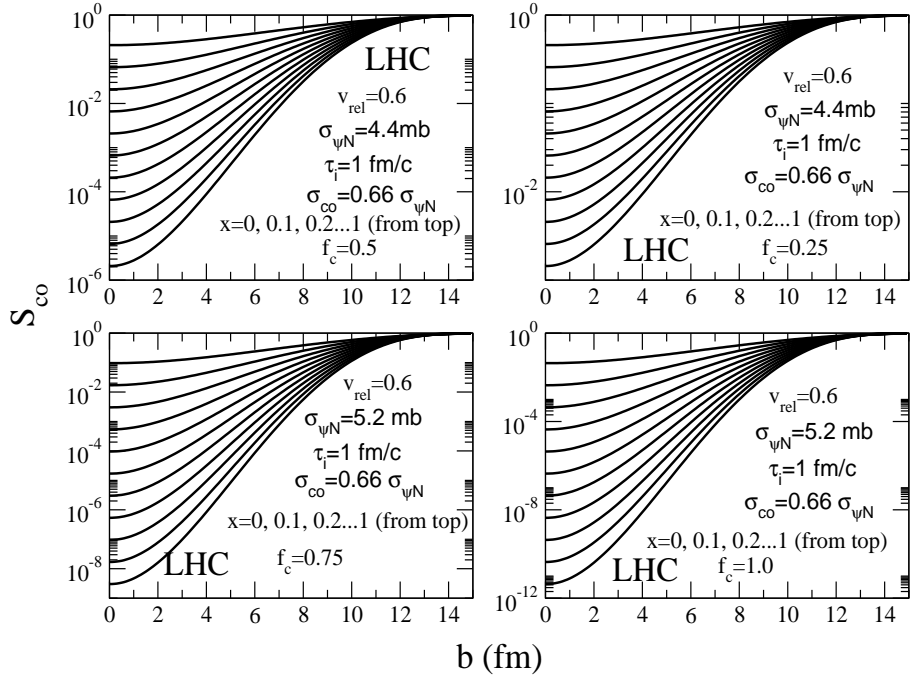
\caption{Comoving survival probability in the scenario II as
a function of impact parameter.
}
\label{fig3}
\eef

Now let us turn to the lepton pair yield from $J/\psi$ decay. 
In fig.~(\ref{fig4}) we compare the lepton pair yield from $Pb-Pb$
collisions at $\sqrt{s} = 6000$ GeV for 0 - 10 \% centrality.
We have chosen different combinations of $x$ and $f_c$ to show the
sensitivity of the results in second scenario of comoving suppression.
However, we have
checked that for very small values of $x$ and $f_c$ the three scenarios
for comoving suppression coincide. But for higher value of both $x$ and
$f_c$ these are quite different. Past (SPS) and present (RHIC) experiments
suggest that the value of $x$ may not be very high even at LHC energies. 
To see the  effects of $x$ and $f_c$ on the dilepton yield we show
in fig.~(\ref{fig4}) the invariant mass distribution for various combinations
of these two parameters in the invariant mass range 
3 GeV$\,\leq M\,\leq\,$ 3.2 GeV. We notice that for the 
combination in which both
$x$ and $f_c$ are large, the yield is less, indicating more suppression
as the beam energy increases (since in that case $x$ 
will be different from zero). 

\bef
\centerline{\psfig{figure=fig4.eps,height=10cm,width=12cm}}
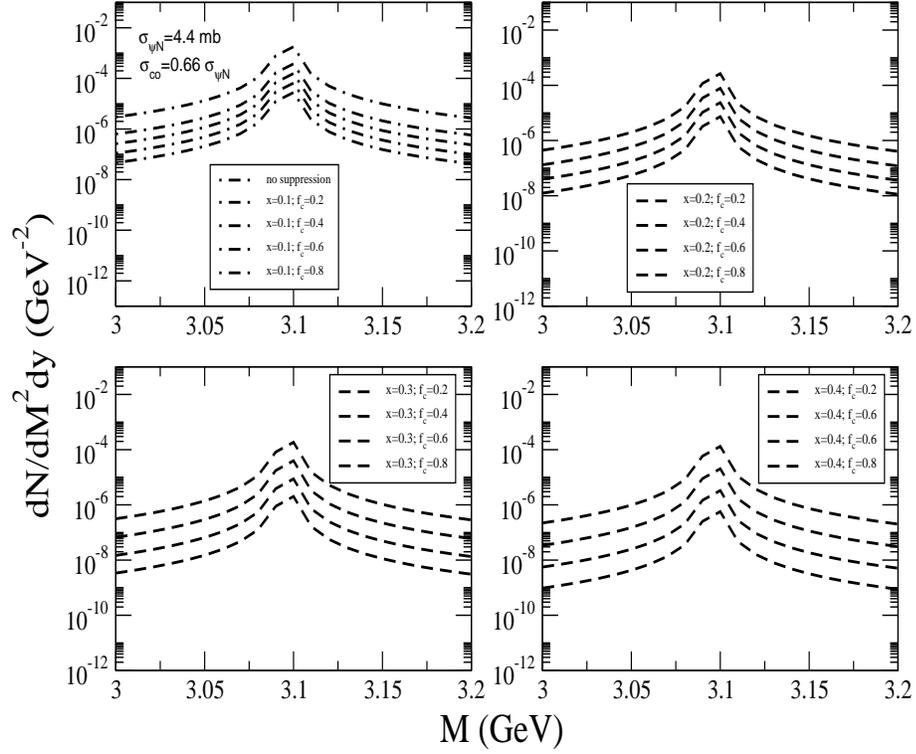
\caption{Dilepton yield in $Pb-Pb$ collisions at $\sqrt{s}
=$ 6000 GeV, for various combinations of ($x, f_c$) in the second case
of comoving absorption. 
}
\label{fig4}
\eef

Now we  compare the hard contribution with the
thermal one in fig.~(\ref{fig5}). For the
thermal part we use simple Bjorken cooling law.
We assume a hadronic matter initial state with initial temperature
of the order of 300 MeV at proper time $\tau_i$ = 1 fm. 
We realise that at $T$ = 300 MeV the hadronic matter the hadronic matter
may dissolve to QGP, however, we take such a high temperature
to show that the hard contribution is 
larger than the thermal even for such a large initial temperature.
For the hard production we choose a
centrality cut of 10 \%. For nuclear shadowing EKS~\cite{eks} 
parametrization has
been used together with CTEQ PDF~\cite{pdf}. 
We have included a $K$ factor $\sim$ 2.5 to account for higher order 
processes.  

\bef
\centerline{\psfig{figure=fig5.eps,height=7cm,width=8cm}}
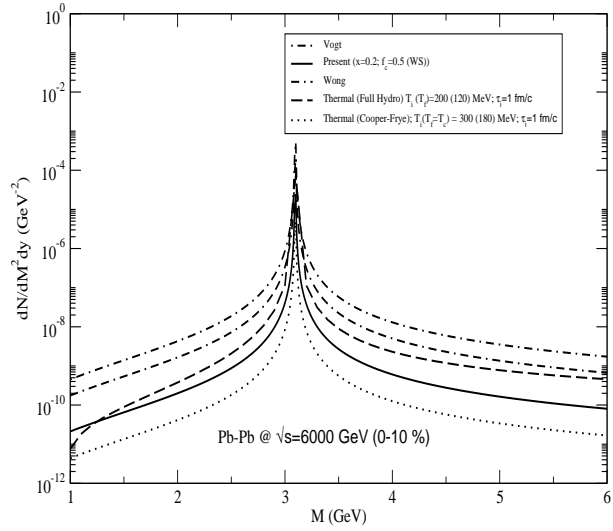
\caption{Dilepton yield in $Pb-Pb$ collisions at $\sqrt{s}
=$ 6000 GeV, $x = 0.2$ and $f_c = 0.5$ in different comoving scenarios
. Thermal contributions are
also included. 
}
\label{fig5}
\eef

As the beam energy increases the shadowing effect in hard scattering processes
becomes important. We have checked that at RHIC energies the lepton pair
yields with and without shadowing are not very different. However, there
is a substantial difference in the yields at LHC energies 
(see fig.~(\ref{fig5b})).

\bef
\centerline{\psfig{figure=fig5a.eps,height=8cm,width=8cm}}
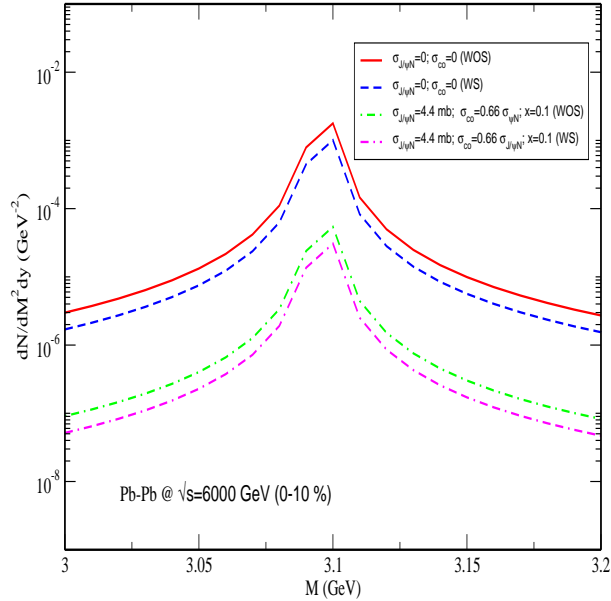
\caption{Dilepton yield in $Pb-Pb$ collisions at $\sqrt{s}
=$  6000 GeV, with (WS) and without (WOS) shadowing for two values 
of absorption cross-sections.}
\label{fig5b}
\eef

The effects of different comoving scenarios on the  
dilepton yield  is clearly seen in fig.~(\ref{fig6}) in the
invariant mass window 3 GeV$\,\leq\,M\,\leq\,$3.2 GeV. We see
that the second scenario of the comoving absorption gives the
maximum suppression for the given set of parameters. Even, the thermal
contribution is larger than the hard contribution.
The yield of lepton pairs (hard part) from the experiment at LHC
energies might lie between these yields.

\bef
\centerline{\psfig{figure=fig6.eps,height=8cm,width=8cm}}
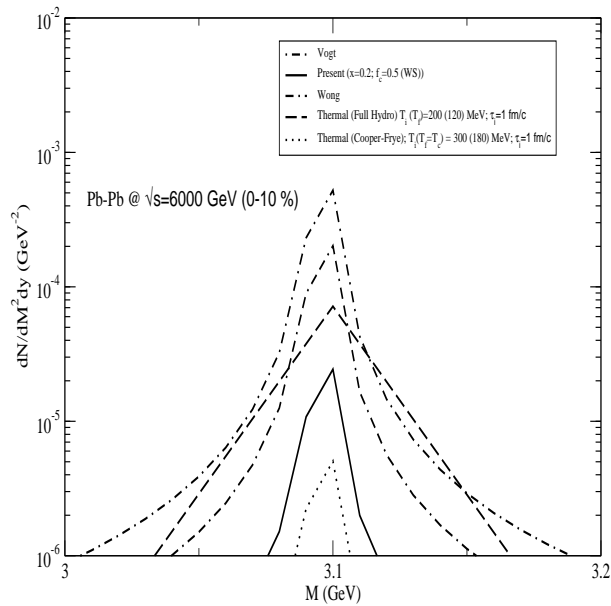
\caption{Same as fig.~(\protect\ref{fig5}) with 3 GeV$\,\leq\,M\,\leq\,$3.2 GeV.
}
\label{fig6}
\eef

\section*{IV. Summary and discussions}
In this work we have calculated the lepton pairs yield from thermal and
hard $J/\psi$ decays.
The effects of the various comover scenarios on 
the survival probability of the heavy quarkonia,
nuclear shadowing on the production cross sections of
the heavy quarks and the centrality of the collisions have  
been considered.  The effects of shadowing and comover  
suppressions on the dilepton spectra resulting from  the decays
of $J/\psi$ is found to be large. 
The dilepton yield from the  thermal $J/\psi$  is found
to be smaller than the contributions from hard $J/\psi$.

The fraction of the comover is treated as a parameter here
and  the lepton pair yields from the heavy quarkonia 
have been evaluated for various values of this parameter.
The increase of $J/\psi$ yield from the decays of 
higher charmonium states have been neglected here,
however the yield with the inclusion of such  processes
may be realized within the parameters ($x$, the fraction of hard component
and $f_c$, the fraction of the comoving hadrons) range considered here.

{\bf Acknowledgment:} One of us (P.R) is grateful to Theory Division, CERN 
PH-TH for hospitality where part of this work was performed.

\end{document}